# Data Quality Taxonomy for Data Monetization

Eduardo Vyhmeister[1], Bastien Pietropoli[1], Andrea Visentin[1]
[1] University College Cork, Cork, Ireland

**Abstract**

This chapter presents a comprehensive taxonomy for assessing data quality in the context of data monetisation, developed through a systematic literature review. Organising over one hundred metrics and Key Performance Indicators (KPIs) into four subclusters (Fundamental, Contextual, Resolution, and Specialised) within the Balanced Scorecard (BSC) framework, the taxonomy integrates both universal and domain-specific quality dimensions. By positioning data quality as a strategic connector across the BSC's Financial, Customer, Internal Processes, and Learning & Growth perspectives, it demonstrates how quality metrics underpin valuation accuracy, customer trust, operational efficiency, and innovation capacity. The framework's interconnected "metrics layer" ensures that improvements in one dimension cascade into others, maximising strategic impact. This holistic approach bridges the gap between granular technical assessment and high-level decision-making, offering practitioners, data stewards, and strategists a scalable, evidence-based reference for aligning data quality management with sustainable value creation.

taxonomy, data quality, balanced scorecard, metrics, KPIs

## 1 Introduction

Over the past decade, the rapid digitisation of business processes and the pervasive adoption of digital technologies have elevated data to the status of a core organisational asset [1, 2, 3]. Companies now allocate substantial resources to collecting, storing, and analysing data, recognising that information of high quality can be a decisive factor in securing competitive advantage. The economic importance of this asset is reflected in its market value, which has reached extraordinary levels; estimated in 2023 at roughly €350 billion in the United States, €82 billion in the European Union, €53 billion in Japan, and €50 billion in China. Numbers that are expected to keep climbing [4]. Yet, despite these impressive valuations, many organisations still find it difficult to quantify the actual economic return generated when data is embedded into their products, services, or decision-making processes [3, 5].

Although widely discussed, data monetisation remains an emerging research field with no single agreed definition and a variety of interpretations [1, 3, 6, 7]. In practice, three broad forms are typically identified. Internal monetisation focuses on improving internal operations and decision-making to deliver tangible benefits, such as cost reductions or efficiency gains. Indirect monetisation enhances the value of products and services through data-enabled features (e.g. instance, recommendation engines, without selling the underlying data). Direct monetisation, by contrast, involves explicitly commercialising datasets, such as through brokers or marketplaces. The related notion of data valuation concerns estimating the worth of a dataset, whether in financial or strategic terms. Because value is context-dependent and often subjective [5], and because monetisation strategies vary widely, determining a standardised "fair" value remains an elusive goal.

Regardless of the monetisation pathway, robust metrics and Key Performance Indicators (KPIs) are essential for evaluating data assets. Traditional data quality metrics, such as accuracy, completeness, and timeliness, are well established in both research and practice [8, 9, 10]. However, quality alone does not capture the multifaceted nature of data value. Other measures, including acquisition and storage costs, accessibility, recoverability after failure, fairness, and bias levels, may be equally relevant depending on the stakeholders involved and the monetisation strategy adopted.

This chapter pursues two main aims. First, it synthesises a wide range of approaches, indicators, and measurement techniques for assessing the value of data, with emphasis on those relevant to monetisation in contemporary and future market settings. Second, it introduces a comprehensive taxonomy of these metrics and KPIs, structured using the Balanced Scorecard (BSC) framework [11]. By aligning financial, customer, internal process, and learning-and-growth perspectives, this taxonomy offers a coherent reference for practitioners and strategists seeking to evaluate and monetise data assets in ways that are both systematic and strategically aligned.

## 2 Definitions and Background

In this section, we present key definitions and background required to understand this study. In particular, because of its connection to DATAMITE, a European research project dedicated to data monetization [12].

## 3 Systematic Literature Review and a New Taxonomy

To develop a robust and comprehensive foundation for the taxonomy and KPI catalogue presented in this work, a systematic literature review (SLR) was conducted following established methodological guidelines [13]. The objective of the SLR was to identify, evaluate, and synthesise existing research and industry practices related to data valuation, quality metrics, and their integration into business-oriented frameworks such as the Balanced Scorecard.

The process began with the definition of clear research questions addressing both conceptual and operational aspects of dataset valuation and its quality dimensions. From there, search strategies were developed to query leading scientific databases and relevant industry repositories, ensuring that both academic and practical perspectives were captured. Search terms were structured around core themes such as data valuation methods, data quality metrics, KPIs for data assets, and data monetisation frameworks.

A multi-stage filtering process was applied to the initial pool of results. First, titles and abstracts were screened to eliminate works outside the scope of dataset valuation and quality assessment. Second, full-text reviews were performed to confirm methodological relevance and empirical grounding. Third, reference snowballing was used to identify additional works cited in key sources. This rigorous process ensured coverage of both foundational works and the most recent developments in the field.

To enable traceability and reproducibility, the inclusion and exclusion criteria, as well as the search strings and databases consulted, were documented in detail. The final corpus of studies was systematically coded and analysed, allowing the extraction of quality metrics, valuation approaches, and linkage patterns with strategic business objectives. These findings were then synthesised into the taxonomy structure, the KPI catalogue, and the mapping to the BSC perspectives presented in this book chapter.

The main results of the SLR, including the complete list of identified metrics, their definitions, associated KPIs, and their classification into taxonomy clusters, are available in the DATAMITE Deliverable D4.2: Basic Computation Approach and Structure KPI Catalogue for the Valuation of Datasets [1]. This deliverable provides the extended tables, methodological appendices, and supporting analysis that underpin the condensed presentation offered in this chapter. Readers seeking the full technical specifications are encouraged to consult D4.2 for a more exhaustive treatment of the literature review outcomes.

## 3.1 Data Quality Taxonomy

The High-Level Taxonomy (Figure 1) structures the strategic evaluation of datasets by aligning data valuation techniques and performance metrics with the four perspectives of the Balanced Scorecard (BSC): *Financial*, *Customer*, *Internal Processes*, and *Learning & Growth*. At the core of the *Internal Processes* perspective lies the *Data Quality* component, which functions as both a standalone dimension of value and a critical enabler for every other cluster in the taxonomy.

*Data Quality* is further organised into four subclusters, each capturing distinct aspects of how well data meets technical, operational, and strategic requirements:

- *Fundamental* quality ensures that datasets are accurate, consistent, and valid, forming the non-negotiable baseline for any valuation or monetization activity.

- *Contextual* quality adapts this baseline to domain-specific requirements such as stability, representativeness, and containment.

---

[1]https://datamite-horizon.eu/deliverables/

- *Resolution* addresses temporal properties such as timeliness and currency, which are crucial for markets and operational contexts where the value of data degrades rapidly.

- *Specialized* quality encompasses usability, clarity, and plausibility, ensuring that data assets are interpretable, fit-for-purpose, and actionable by end users.

Within the taxonomy, the importance of Data Quality is twofold:

**Direct Contribution to Data Value:** The Quality & Utility clusters, of which Data Quality is a core part, feeds directly into the calculation of a dataset's value, influencing market price, monetisation potential, and prioritisation for internal use. Without strong Data Quality, even datasets with high potential utility may fail to generate economic returns or meet strategic objectives.

**Cross-Perspective Enabler:** Although positioned under Internal Processes, Data Quality exerts influence across all BSC perspectives. High-quality datasets strengthen the Financial perspective by reducing valuation risk and ensuring defensible pricing. They improve the Customer perspective by delivering accurate, relevant, and trustworthy insights, directly impacting customer satisfaction and retention. In the Learning & Growth perspective, quality ensures that innovation initiatives and analytical models are built on reliable foundations, reducing rework and accelerating development cycles.

In essence, Data Quality acts as the structural integrity of the taxonomy: it is the dimension that transforms raw information into a dependable strategic asset. Without it, the effectiveness of governance, compliance, operational efficiency, and even advanced valuation techniques is severely diminished. By embedding Data Quality into the heart of the taxonomy, the framework ensures that all strategic objectives are supported by data that is not only available but also fit for its intended purpose.

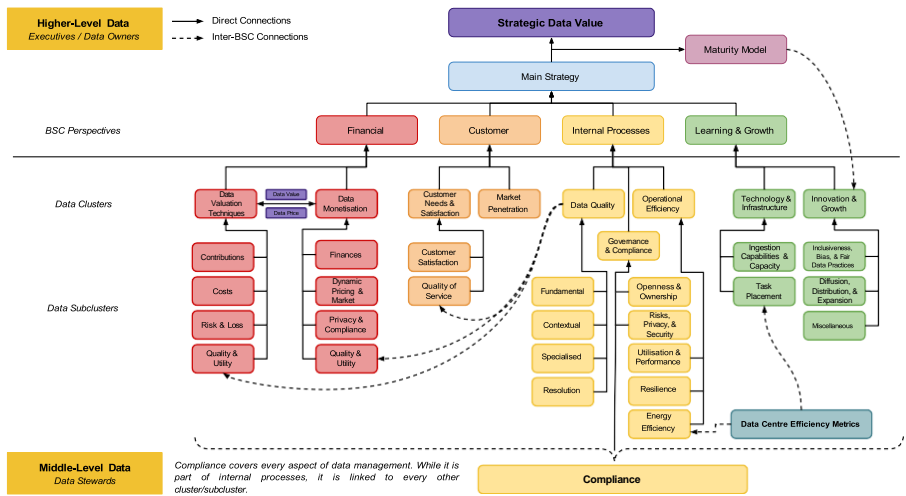

Figure 1: Higher Level Taxonomy Diagram

The Data Quality cluster in the Internal Processes perspective is further detailed in Figure 2. It provides a detailed view of how the four subclusters are interwoven through a shared metrics layer. Rather than existing as isolated domains, these subclusters interact dynamically, with metrics acting as bridges that create continuity and integration across the entire quality dimension.

At the structural level, the figure shows that the metrics layer operates both vertically and horizontally. Vertically, it aligns each subcluster's conceptual attributes with measurable indicators, ensuring that every aspect of quality can be operationalised. Horizontally, it allows for cross-subcluster interaction, where a single metric can inform multiple dimensions of quality. For example, a timeliness-related measure may originate in the Resolution subcluster but also influence Contextual relevance and Specialised usability.

This interconnectivity is not accidental. It reflects the reality that quality attributes are interdependent. The figure illustrates how metrics provide the common language that allows fundamental principles like integrity and consistency to inform Contextual assessments such as volatility or representativeness. Similarly, metrics derived from Resolution (e.g., update frequency) often underpin Specialized characteristics like clarity or usability, as data currency impacts how users perceive accessibility and trustworthiness.

By mapping these relationships visually, the figure makes clear that metrics act as integrative nodes. They prevent fragmentation of quality assessment by ensuring that evaluation in one subcluster has a traceable influence on others.

This is particularly relevant in the Balanced Scorecard context, where isolated improvement in one process area may have little strategic impact unless it cascades into others. The metrics layer ensures such a cascade happens naturally.

Another key implication of this interconnected structure is measurement efficiency. The same metric can serve multiple assessment purposes, reducing redundancy in monitoring systems. For example, a measure of data completeness, anchored in the Fundamental subcluster, can simultaneously serve as an indicator of operational readiness in the Contextual subcluster and contribute to clarity in the Specialised subcluster.

The metrics layer also supports multi-perspective alignment within the BSC framework. Because metrics inherently cross subcluster boundaries, they facilitate the translation of Internal Processes insights into Financial, Customer, and Learning & Growth impacts. This is achieved by tracking quality at multiple points of influence, ensuring that improvements in one area produce measurable benefits in others.

In sum, the figure reveals that the metrics layer is not merely a measurement block. It is the connective tissue of the Data Quality dimension. By linking subclusters into a coherent, mutually reinforcing network, it ensures that quality management is holistic, strategically aligned, and capable of driving value across the entire taxonomy.

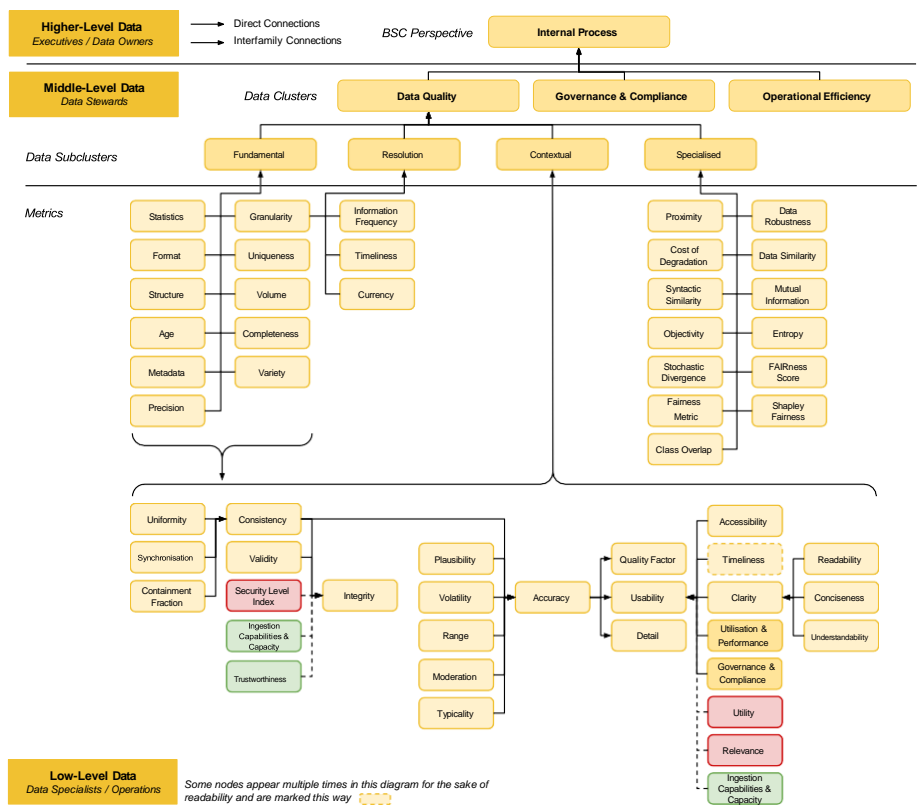

Figure 2: Lower Level Taxonomy Diagram - Internal Process & Data Quality

It is important to note that the terminology and definitions associated with data quality dimensions vary considerably across the literature. As highlighted by [14], there are at least 76 distinct dimensions in use. Furthermore, different frameworks to cluster quality dimensions have alse been performed, though disconnected from an enterprise perspective [15]. Short definitions for each of these dimensions are provided in the DATAMITE product [2].

This diversity of terminology reveals overlapping concepts (e.g. Credibility and Trustworthiness), dimensions not strictly related to quality (e.g. Privacy), and metrics that are inconsistently aggregated (e.g. Format Precision). Such variability underscores the need for the present taxonomy, which reorganises these measures according to their relative importance from a business-driven perspective.

### 3.1.1 Fundamental

Within the fundamental dimension, a number of metrics can be employed to characterise and summarise the contents of a dataset. Basic statistical indicators, such as average, standard deviation, minimum, and maximum, are part of this group, although they are not discussed in detail here. In Figure 2, they appear collectively under Statistics.

**Age** measures the time elapsed since a data item was generated, updated, or last accessed [16, 17]. It is an important indicator of Timeliness and Relevance across many use cases. Depending on its application, Age can be calculated in two distinct ways:

- Static Age – the interval since an entire dataset or information record was first created. This form often informs archival or tiered storage decisions, where older datasets are assigned lower-priority storage due to diminished utility.

- Dynamic Age – the interval since the most recent update to the dataset was received by a user or system, serving as a measure of data freshness.

**Granularity** (also known as Abundance or Data Frequency) denotes the level of detail at which data is collected, stored, and analyzed, irrespective of refresh cycles. It determines the smallest unit of information available and shapes how precisely data can be interpreted. Fine-grained data offers more specific insights but demands greater storage, processing capacity, and computational effort [18]. Conversely, coarse granularity sacrifices detail in favor of efficiency and broader trend detection [19]. The optimal level is dictated by analytical objectives, dataset characteristics, and the trade-off between precision and manageability [20].

Granularity is central to understanding the temporal dynamics of data flow, especially in domains such as real-time analytics, IoT networks, or streaming platforms. High-frequency scenarios include financial market feeds with millisecond updates, or continuous sensor outputs from IoT devices [18, 19, 20, 21].

---

[2]https://datamite-horizon.eu/deliverables/

While sometimes linked to Velocity [22], the two differ: velocity relates to the rate of data generation and movement, focusing on timeliness and real-time performance, whereas data frequency concerns the temporal resolution of individual events. Consequently, velocity aligns more closely with the Technology and Infrastructure cluster.

**Precision** describes the fineness with which data values are recorded and represented, affecting the reliability of analytical outcomes [14, 23, 24, 25]. It can be divided into three facets:

- Numerical precision – the number of significant digits or decimal places (e.g., 23.456 is more precise than 23.5).

- Consistency – the degree to which repeated measurements produce the same result under identical conditions.

- Repeatability – the ability to obtain the same measurements over multiple trials.

Precision is not confined to numeric data; for example, "Scarlet," "Azure," and "Emerald" convey more precise categorization than broad labels like "Red," "Blue," or "Green."

**Uniqueness** (or Redundancy in service contexts) reflects the extent to which dataset entries are distinct and duplicate-free. It ensures each entity is represented only once, safeguarding data integrity and supporting consistent decision-making [14, 26].

**Variety** (also referred to as Multifacetedness) addresses the diversity of data types, formats, and sources, including structured, semi-structured, and unstructured content. It captures the heterogeneity of modern data ecosystems and the complexity of integrating features from multiple origins. Greater variety often increases analytical potential but also raises computational demands [27].

**Volume** (alternatively labeled quantity, entries, total data amount, number of data sets available or information quantity) represents the amount of data available for processing and analysis [17, 18, 27, 28, 29, 30]. It may refer to the total number of records, points, or entries in a dataset and influences:

- Statistical significance, by reducing sampling error.

- Pattern recognition, by enabling trend and correlation discovery.

- Machine learning performance, by providing sufficient training data.

Adequate dataset size ensures coverage for its intended applications, while access to multiple datasets increases flexibility and reuse potential [28, 29].

**Metadata**, closely related to Profiling, captures descriptive information about data structure, provenance, content, and context. It underpins dataset interpretation, discovery, and usability. Quality frameworks often evaluate metadata in terms of semantic consistency, vocabulary usage, and currency [31, 32]. From a metric perspective, metadata can be treated as a binary indicator of whether relevant descriptive fields exist. Its functions include:

- Discoverability making datasets searchable and contextually interpretable (e.g., through format specifications).

- Quality assessment – evaluating dataset accuracy, completeness, and trustworthiness.

- Context provision – detailing origins, scope, and intended use for alignment with specific applications.

**Format** (also known as Format Compliance, Codification, Conformity, or Available Formats) defines both the structural arrangement of data and the proportion of entries conforming to a required pattern. Compliance ensures datasets meet expected schema or structural requirements (e.g., for geospatial coordinates) [31, 25, 21, 33].

**Structure** refers to the organization of data, whether structured, semi-structured, or unstructured. It shapes usability and determines processing complexity. Syntactic and semantic rules ensure consistency, interoperability, and analytical readiness. Well-structured datasets yield higher value, while poorly structured ones demand costly pre-processing [18, 14, 33].

**Completeness** (sometimes termed Appropriate Amount of Data) measures the degree to which all required and expected values are present. This widely discussed dimension [26, 14, 17, 31, 25, 33, 28, 23, 22] can also be linked to metadata to verify whether key fields are populated, or to assess coverage relative to an intended application.

### 3.1.2 Contextual

These are metrics that evaluate data quality relative to a specific application, use case, or domain context. In other words, these metrics confirm that the quality of data is not absolute but depends on who is using it, why, and under what conditions.

**Range** represents the share of data values falling within predetermined minimum and maximum limits. It serves as an indicator of whether data lies within the expected operational boundaries, directly influencing the *Utility* dimension of data quality. Particularly for numerical attributes, maintaining values within domain-specific or statistically derived thresholds (e.g., limits based on extrema or quartile boundaries) is essential for ensuring validity [34, 35].

**Moderation and Typicality** – When relevant, Moderation quantifies the proportion of data contained within a specific confidence interval. For example, under a normal distribution, roughly 99.7 % of data points are expected to lie within three standard deviations of the mean. This measure reflects the stability and reliability of a dataset by assessing adherence to high-confidence expectations [34]. Typicality, in contrast, measures the extent to which a single data item matches established or "usual" patterns in the dataset. It highlights how representative a value is compared to expected norms, and aids in identifying anomalies or unusual events, particularly in dynamic datasets where deviations may hold greater relevance [34].

**Volatility** describes how frequently data values change over time and, in certain contexts such as finance, also captures the magnitude of those changes [14]. The method of estimation depends on the intended purpose. Regardless of the approach, volatility reflects a variable's instability and update frequency, with the weight assigned to different timeframes depending on the feature's characteristics. This context-dependent metric can also be calculated directly without requiring foundational quality measures.

**Consistency** is closely linked to Veracity and Reliability [27]. Consistency measures the alignment of data with established rules. These rules may be structural, such as relational database constraints, or derived from metadata and organizational standards. Following [36], intra-consistency evaluates whether a tuple satisfies or violates any rule from a set of association rules. In a broader sense, inter-consistency concerns the uniformity of information across datasets, systems, and applications, ensuring accuracy, coherence, and absence of contradictions.

Key elements of consistency, include [18, 14, 35, 31, 28, 34, 23]:

- Integrity – Data remains accurate and dependable, producing the same outcome regardless of access point or processing route.

- Uniformity – Formats, naming conventions, measurement units, and data types are standardized.

- Synchronization – Updates are propagated across all connected systems to avoid mismatches.

- Validation Rules – Constraints enforce correct types, ranges, and relationships.

- Error Prevention – Processes are in place to detect and correct conflicting entries.

- Data Governance – Policies establish and maintain organizational standards.

**Containment Fraction** measures how much of one dataset exists within another. This information is particularly valuable in distributed systems for evaluating consistency, detecting redundancy, and optimizing storage [105].

**Integrity**, which is defined as the correctness, reliability, and adherence of data to standards or rules, it guarantees completeness, authenticity, and suitability for its intended use [26, 14, 35, 21, 33, 29, 16, 23]. Consistency is one aspect of integrity, while Reliability, heavily dependent on integrity, is often treated alongside it.

**Uniformity** reinforces consistency by enforcing identical representation formats (e.g., date patterns, numeric formats), without evaluating logical correctness, which is the focus of Validity.

**Validity** ensures that each individual value adheres to logical and operational rules, for instance, that age or costs are positive numbers [35, 29, 23].

As an analogy: in a puzzle set, integrity ensures all original pieces are present and undamaged; validity confirms each piece is correctly shaped; consistency ensures the pieces fit together without conflict; uniformity guarantees all pieces share the same material and finish.

**Accuracy** measures how closely data reflects the true or intended state of the world. It encompasses not only correctness but also suitability for decision-making and analysis [24, 26, 14, 35, 25, 21, 33, 29, 28, 22]. Metrics contributing to accuracy include range, consistency, moderation, and typicality [34]. Acceptable accuracy thresholds are application-specific, with small deviations permissible depending on context.

**Quality Factor**, as specified in [24], estimates how quality supports business innovation, combining accuracy with frequency. The frequency element here corresponds to Information Frequency (IF), discussed under resolution metrics.

**Detail and Plausibility** – Detail assesses whether information is recorded with sufficient precision, linking accuracy to data valuation [26]. Plausibility gauges whether data aligns with real-world knowledge, a definition extended in the ECB framework to include outlier detection [26, 14, 32, 33, 28, 23]. As a KPI, plausibility may integrate range, consistency, domain-specific rule adherence, and anomaly detection.

**Usability** is another relevant, widely discussed, and with considerable context-dependent connotation metric. Usability can be defined as a hierarchical high metric (i.e. a combination of several components). In fact, Usability according to ISO 9241: 11 (2018), is a benchmarking tool that can be used to determine the extent to which a system, product, or service can be used by specific users to achieve the goals determined by the effectiveness, efficiency, and satisfaction of its users." [32]. Building on the definition of variable indexes from [32] (see Table 1) and insights from other sources [14, 17, 25, 21, 33, 23] that explore the concept of Usability, we have expanded the term to mean: "The ease and efficiency with which quality data, defined by Data-Value, Accuracy, Integrity, and Completeness, can be accessed (through Communication, Accessibility, and Timeliness), understood (Clarity), and effectively used (Ease-of-use, linked to Operational Efficiency, Openness & Performance, Relevance, or Utility) by users to complete specific tasks". Importantly, not all these components are relevant at the same time and others could be further linked since they depend on the strategies involved. Usability ensures that data not only meets Technical Standards (highly related to Quality and Governance) but also aligns with the practical needs of users, enabling them to extract meaningful insights, make informed decisions, and complete tasks efficiently [33, 23]. Based on the previous description, Usability can be treated as a KPI, rather than a metric, in which a combination of metrics (or other KPIs) are used in its calculation.

**Clarity** is a composite KPI reflecting how unambiguously and readably information is presented, improving interpretability and usability [26, 14, 31, 32, 28, 23]. Since clarity is a combination of factors, it can be treated as a KPI where the values are agglomerated and weighted. Clarity serves as an integrative measure reflecting several metrics (see the following ones), emphasising the presentation of high-quality, actionable, and user-centred information across diverse applications, from open data platforms to governance and compliance systems.

**Conciseness** refers to the principle of presenting data or information in a way that is both brief and clear, without unnecessary detail or redundancy. As a metric, Conciseness helps measure clarity. Furthermore, it can be used as an efficient representation of data quality, fostering minimisation of redundancy while preserving functionality. Conciseness (or Unambiguity) ensures that information is free from unnecessary complexity and includes only relevant data. It also refers to the removal of ambiguity. For example, unambiguous schemas and standardised metadata and presentation of data play key roles in improving data quality and ensuring consistency across systems [26, 14].

**Understandability**, or Ease of Understanding, refers to how easily users can interpret information. Explicit references to understandable data include clear field names, precise definitions, and avoidance of ambiguous units so that users can comprehend datasets without extensive technical expertise [14, 31, 32]. Readability enhances clarity by structuring information for intuitive use. This includes proper formatting, well-designed user interfaces, and visual representations, making it easier for users to interact with and extract meaning from data [37, 38].

### 3.1.3 Resolution

As previously described, within this group we agglomerate metrics that are time, dynamic, or strongly granulometry-dependent. Within this group, Timeliness (Data Freshness) and Currency are considerably recognized [26, 14, 35, 25, 21, 33, 29, 28, 34, 16, 23, 22].

**Timeliness and Currency** - Timeliness is a measurement of the delay between an event occurring and the data being available to the business, therefore is strongly dependent on Age. On the other hand, Currency is whether the data has lost its value due to its processing, modification or elapsed time and thus can be linked to change in data value with respect to time change or an event, thus it can be considered to be dependent on data valuation techniques.

**Information Frequency (IF)** is defined as the rate at which information is updated, accessed, or utilised within a system or process. It reflects the temporal characteristics of information flow and interaction, ensuring that information remains relevant and aligned with the operational or decision-making needs. This metric is linked to operational efficiencies, but given its relative connection to quality concepts, it has been incorporated into this cluster. IF is integrated into multiple dimensions, including:

- Raw Frequency Component (IFr): A baseline measure of how often information or data becomes available.

- True Frequency Component (IF): A refined measure of frequency adjusted for external factors influencing the information flow.

- Frequency Tolerance (FT): The range of acceptable frequency deviations that still supports effective decision-making.

- Node Frequency Requirement (FN): The specific frequency threshold necessary to satisfy the operational or decision-making needs of a given system.

This aggregation captures both quantitative and qualitative aspects of frequency, ensuring that information meets the availability requirements of the baseline and is provided with the precision and timeliness necessary for optimised decision-making processes [24]. Even though each of these frequencies could be defined as individual metrics (or linked to others described in this work), the direct connection to IF made us establish it separately and be considered only as part of the calculation of IF.

### 3.1.4 Specialised

Specialized metrics are domain-specific or task-specific measures of data quality that go beyond general-purpose dimensions (like accuracy or completeness, or even context wise metrics). They are tailored to particular industries, data types, or applications, often to meet regulatory, operational, or scientific requirements.

**Entropy** and other advanced metrics such as Mutual Information, and Shapley Values expressions (e.g. Shapley Fairness and Shapley Robust) provide deeper insights into the structure and content of datasets. Entropy is a measure of uncertainty, randomness, or information content within a system, dataset, or process. It quantifies the degree of unpredictability or heterogeneity in data and is used to assess information richness, uncertainty, or the effectiveness of data representation. Commonly rooted in information theory, entropy plays a crucial role in evaluating data quality, value, and informativeness [30, 22]. Although Entropy has extensively been used for valuation, its incorporation as a metric is rooted through its connection to quality and thus, has been established in this work as a Quality metric connected to valuation techniques. The various forms of Entropy (as clustered in this work under the same metric name), including Shannon's Entropy, Heterogeneity, Information Entropy, Additional Information Value (AIV), Joint Entropy, Individual Entropy, Information Score metric are context dependent and thus the decision of which one to implement depends on the strategy involved within business models.

**Mutual Information**, as defined in information theory, focuses on quantifying the amount of information shared between two random variables. It measures how much knowing one variable reduces uncertainty about the other. It is expressed mathematically as the Kullback-Leibler divergence between their joint probability distribution and the product of their marginal distributions. This makes mutual information a powerful tool for assessing dependencies or relationships between datasets or variables, often applied in data science, machine learning, and signal processing [16, 42]. Mutual information plays a critical role in evaluating Relevance, Consistency, and Completeness. For instance, low mutual information may highlight a lack of coherence in the data, pointing to potential gaps or errors. Furthermore, it serves as a key metric in feature selection and dimensionality reduction, ensuring that only the most relevant and informative attributes are preserved, thereby enhancing the quality of data used in downstream processes [16, 42].

**Shapley Fairness and Fairness Metric**, as covered in [41], builds on the concept of Shapley Value. The latter quantifies the contribution of individual participants in a cooperative system. It ensures that the allocation of rewards or resources respects specific fairness principles, such as balance (distributing the total value amongst participants), symmetry (equal rewards for equal contributions), zero element (no reward for no contribution), and additivity (consistent allocation across combined tasks). This framework is particularly relevant in data-centric contexts as it ties closely to data quality. Shapley Values help assess the importance of individual data points or datasets to the performance of predictive models, reflecting the impact of high-quality data. They can also highlight redundancy or noise by identifying data that adds little value, encouraging improvements in data quality. Based on the content of [43, 44], the Fairness Metric (linked to Shapley fairness) is a measure designed to quantify the equitable allocation of resources, contributions, or compliance with specific principles in diverse contexts. For example, the Fairness Metric is used in a Weight-based Fair Share Algorithm for allocating cache space amongst virtual machines (VMs) [43]. It ensures proportional allocation based on pre-assigned weights.

**FAIRness Score**, not to be confused with the previous metrics, evaluates dataset's compliance with the FAIR principles (Findable, Accessible, Interoperable, and Reusable) [39]. Automated tools like CkanFAIR compute this score to assess how well datasets meet these criteria, aiming to improve data quality and promote effective sharing.

**Data Similarity**, named in this work, encapsulates measures such as Euclidean Distance, Projection Similarity, Similarity Score, Cosine Similarity, Average Distance, Kolmogorov-Smirnov (KS), Jaccard Similarity, Mann-Whitney (MW), Mood's Median (MD), and Levene (LE)) [16, 45, 30]. These metrics are used to evaluate similarities or differences between sets, vectors and properties. For example, Jaccard Similarity is ideal for determining the overlap between datasets or sets.

**Syntactic Similarity** (similarly linked to Levenshtein Distance, edit distance, cosine similarity, Q-gram distance, semantic similarity, Jaccard coefficient, MinHash-Based Distance, Overlap Set Similarity, String-based Measures) focuses on assessing how similar data values are in terms of their syntax and are widely used in text analyses [30]

**Stochastic Divergence** (named by us) corresponds to metrics that measure the similarity between probability distribution (e.g. T-test scores, Identity-based Exact Match, Jensen-Shannon Divergence, Wasserstein distance).

**Objectivity** is a measure of the degree to which data or a data source is believed to be free from biases, ensuring that the information presented is impartial and unaffected by subjective influences (during collection, evaluation, and use).

It is categorised as an intrinsic or subjective data quality dimension and is critical for assessing the reliability and credibility of data used for decision-making [14, 28, 23]. Since Objectivity is linked to the measure of bias, metrics already known to be used to measure it can be directly linked (e.g. Non-parametric cohort analysis, Statistical Parity, Distributional Skewness, Equalised Odds, and others).

**Cost of Degradation, Information Content (IC), and Proximity:** Cost of Degradation quantifies the loss in data quality resulting from data transformation [46]. IC is related to the data acquisition process. Some data might be very common or predictable, while other data might be rare or surprising. IC gives higher importance to the surprising, less predictable data because it's often more useful for understanding new or important events. Finally, Proximity is described as a factor related to the physical distance of the source of an event. This metric is relevant when there exists a correlation between the quality of the data and the distance of the sensor from the event [16].

**Data Robustness and System Robustness** Robustness refers to the ability of a system, model, or process to remain stable and perform well despite disturbances, faults, or unexpected inputs. It is a measure of how well a system can handle variability or adversity while still functioning as intended. Thus, Robustness for systems can be tracked as a combination of Stability - The ability of a system to maintain functionality despite errors or faults; Resilience to Adversarial Attacks and Change - Particularly in machine learning and Data Management, robustness refers to the ability of a model to resist manipulation by adversarial inputs. Concerning robustness in data, it can be seen as a separate concept not related to statistics. It is related to the concept that data is constructed, acquired, manipulated to survive and function in multiple settings (i.e. resilience to Change but not necessarily to Stability, which is related to the system). Given these two concepts, we have defined two metrics: Data Robustness and System Robustness. For example, in [41] the emphasis is on data robustness in the context of creating a marketplace for data. This ensures that datasets maintain their value and usability across various prediction tasks. Alternatively, in [40], the focus is on system robustness within federated learning architectures for data marketplaces. It discusses designing a robust model aggregation protocol that excludes low-quality or malicious contributions, ensuring inclusiveness and resilience against attacks.

Robustness is normally classified within Data Quality metrics but given its broader perspective and meaning within different domains, in the present work, it has been included in two domains, Data Quality and Operational Efficiency.

### 3.2 Connection to other BSC perspectives

Within the taxonomy, Data Quality is positioned under the Internal Processes perspective of the Balanced Scorecard (BSC). However, its influence extends far beyond its immediate cluster, establishing functional and strategic links with the other three BSC perspectives: Financial, Customer, and Learning & Growth. This interconnection ensures that the measurement and improvement of quality-related metrics do not occur in isolation but contribute directly to broader organisational objectives defined in the BSC structure.

While the subcluster of data quality serve distinct functions in evaluating and maintaining data, its impact extends beyond the Internal Processes perspective, influencing strategic objectives across the Financial, Customer, and Learning & Growth perspectives of the Balanced Scorecard (BSC).

Table 1 summarises the high-level connections between each Data Quality subcluster and the other BSC perspectives, illustrating how quality serves as a strategic connector between technical characteristics and organisational performance.

Table 1: Connections Between Data Quality Subclusters and Other BSC Perspectives

| Data Quality Subcluster (Internal Processes) | Financial Perspective | Customer Perspective | Learning & Growth Perspective |
|---|---|---|---|
| **Fundamental** | Ensures valuation and monetisation models are based on reliable, defensible data, reducing financial risk and enabling stable revenue generation. | Provides consistent, accurate, and relevant outputs that enhance customer satisfaction, trust, and adoption of data-driven services. | Supports robust, error-free datasets for innovation, reducing rework and increasing success rates in R&D and technology projects. |
| **Contextual** | Improves the precision of valuation models by incorporating stability, representativeness, and redundancy measures; informs risk and pricing strategies. | Delivers datasets tailored to specific use cases and market conditions, improving perceived relevance and alignment with customer needs. | Guides system design and innovation processes for dynamic or variable data environments, ensuring scalability and adaptability. |
| **Resolution** | Enhances financial decision-making in time-sensitive markets by maintaining data freshness and relevance, impacting dynamic pricing and valuation accuracy. | Provides up-to-date insights to customers, increasing engagement and decision confidence. | Drives infrastructure requirements for real-time analytics and supports agile innovation cycles dependent on frequent data updates. |
| **Specialised** | Expands monetisation potential by increasing the applicability and market appeal of datasets; supports transparent valuation reporting. | Improves ease of use and interpretability for customers, fostering higher adoption rates and stronger brand positioning. | Accelerates innovation by reducing onboarding time, enabling cross-functional collaboration, and ensuring dependable experimental outcomes. |

### 3.2.1 Financial Perspective

The Financial perspective in the taxonomy comprises clusters such as Data Valuation Techniques and Data Monetisation. Quality indicators, such as accuracy, completeness, consistency, timeliness, can directly feed into these clusters by determining requirements and applicability of the datasets used for valuation and pricing. High-quality data strengthens cost-based, contribution-based, and utility-based valuation methods, thereby improving the reliability of derived KPIs like ROI from data-driven projects or calculated data value (DV). Con-versely, valuation techniques influence how quality is prioritised; for example, in contribution-based models (e.g. Shapley Value), higher-quality inputs increase a dataset's relative contribution and hence its share of monetisation benefits. In dynamic pricing scenarios, maintaining quality reduces the need for corrective adjustments, stabilising revenue flows. Thus, Data Quality acts as both an input factor for financial measurement and a value-protection mechanism in monetisation strategies.

### 3.2.2 Customer Perspective

The Customer perspective includes *Customer Needs & Satisfaction* and *Market Penetration clusters*. Quality metrics from the Internal Processes perspective underpin these by ensuring that data products and insights meet or exceed user expectations. Metrics such as integrity, validity, usability, and clarity influence customer-facing KPIs like satisfaction scores, retention rates, and adoption of new data products. For example, datasets with strong consistency and plausibility support the delivery of accurate, relevant, and timely insights to clients, fostering trust and loyalty. Timely and uniform data presentation improves the perceived value of services, directly impacting market penetration KPIs. Furthermore, volatility and moderation measures can be applied to customer-driven use cases where stability of data is critical for user confidence. In this way, Data Quality sustains the value proposition that drives customer acquisition and retention.

### 3.2.3 Internal Processes Perspective

Within its own BSC perspective, Data Quality interacts closely with other internal clusters, particularly Data Governance & Compliance and Operational Efficiency. Governance metrics, covering compliance with standards, risk reduction, and secure handling, shape the frameworks within which quality is measured and maintained. For instance, governance rules define acceptable ranges, validation procedures, and metadata standards that directly determine quality scores. Operational efficiency metrics, in turn, rely on quality to avoid rework, duplication, and decision delays. High-quality datasets ensure that process efficiency improvements, resource utilisation gains, and resilience measures are sustainable. The link is bidirectional: operational improvements reinforce data quality, while quality improvements enable more efficient processes.

### 3.2.4 Learning & Growth Perspective

The Learning & Growth perspective contains *Technology & Infrastructure* and *Innovation & Growth-Oriented* clusters. Technological capacity, such as ingestion capabilities, processing power, and integration infrastructure, directly affects the ability to sustain key quality metrics like timeliness, uniformity, and accessibility (i.e. inverse direction; from technological capacity to quality). Independent of these type of connections, as seen in the figure, the linkage on direct directional from quality to learning & growth was kept mainly throughout maturity models, that facilitate to understand need and requirements in order to improve the current status of the data approaches. Furthermore, poor-quality data slows innovation cycles and increases the risk of failed initiatives, whereas quality-by-design approaches integrated into innovation pipelines enhance the organisation's adaptability and long-term growth potential.

# 4 Conclusion

This chapter has presented a comprehensive framework for understanding and operationalising data quality in the context of data monetisation. Through a systematic literature review, we consolidated a diverse and often fragmented set of metrics, KPIs, and conceptual dimensions into a coherent taxonomy aligned with the Balanced Scorecard (BSC) framework. By structuring data quality across the Fundamental, Contextual, Resolution, and Specialised subclusters, the taxonomy captures both universal and domain-specific quality requirements, enabling their systematic integration into valuation and monetisation strategies.

A central contribution of this work is the recognition that data quality is not an isolated technical concern but a strategic connector across all BSC perspectives. High-quality datasets strengthen financial outcomes by underpinning reliable valuation and pricing models; they enhance customer satisfaction and trust through relevance, clarity, and usability; they support operational efficiency by reducing rework and ensuring governance compliance; and they enable innovation and technological growth by providing robust, trustworthy foundations for new products and services.

The interlinked metrics layer ensures that improvements in one quality dimension cascade into others, maximising the strategic impact of quality management. This integrated approach allows organisations to balance short-term monetisation goals with long-term value creation, ensuring that data assets remain not only profitable but also sustainable and adaptable in evolving market and regulatory environments.

By embedding quality as a foundational element of data valuation and monetization, this taxonomy offers practitioners a scalable, evidence-based reference for aligning technical measures with business objectives. In doing so, it bridges the gap between granular measurement and strategic decision-making, providing a pathway for organizations to unlock the full potential of their data assets.

# 5 Acknowledgment

This research was partially supported by the EU's Horizon Digital, Industry, and Space program under grant agreement ID 101092989-DATAMITE. Additionally, we acknowledge Science Foundation Ireland under Grant No. 12/RC/2289 for funding the Insight Centre of Data Analytics (which is co-funded under the European Regional Development Fund).